\documentclass[fleqn,12pt,a4paper]{article}

\usepackage[normalem]{ulem}
\usepackage{color}

\usepackage{microtype}
\usepackage{lmodern}
\usepackage[T1]{fontenc}
\usepackage{amssymb}
\usepackage{mathptmx}
\usepackage{graphicx}
\usepackage{url}
\usepackage{hyperref}
\hypersetup{
linktocpage=true,
breaklinks=true,
    frenchlinks=true, %
    unicode=false,          
    pdftitle={Numerical Evaluation of Tensor Feynman Integrals in Euclidean Kinematics},    
    pdfauthor={J. Gluza, K. Kajda, T. Riemann, V. Yundin},     
    pdfsubject={tensor Feynman integrals},   
    colorlinks=false,       
    linkcolor=red,          
    citecolor=green,        
    filecolor=magenta,      
    urlcolor=cyan           
}
\usepackage[all]{hypcap}
\usepackage{amsmath}
\usepackage{pstricks}
\usepackage{pst-node}
\usepackage{a4wide}

\numberwithin{equation}{section}
\numberwithin{figure}{section}
\numberwithin{table}{section}

\usepackage{rotating}

\def \ps {p\hspace{-0.43em}/}

\def \math {\textsf{Mathematica}{}}
\def \ginac {\textsf{Ginac}{}}
\def \ar {\texttt{AMBRE}{}}
\def \mb {\texttt{MB}{}}
\def \mbres {\texttt{MBresolve}{}}
\def \barnesr {\texttt{barnesroutines}{}}
\def \cs {\texttt{CSectors}{}}
\def \secdec {\texttt{sec\-tor\_de\-com\-po\-si\-tion}{}}

\def \mbm {\texttt{MB.m}{}}

\newcommand{\bea}{\begin{eqnarray}}
\newcommand{\eea}{\end{eqnarray}}
\newcommand{\bqa}{\begin{eqnarray}}
\newcommand{\eqa}{\end{eqnarray}}

\newcommand{\nl}{\nonumber \\}
\newcommand{\ep}{\varepsilon}
\newcommand{\eps}{\varepsilon}

\newcommand{\be}{\begin{equation}}
\newcommand{\ee}{\end{equation}}
\newcommand{\ba}{\begin{eqnarray}}
\newcommand{\ea}{\end{eqnarray}}

\begin{document}

\def\sigmap{\sigma^{\prime}}
\def\mup{\mu^{\prime}}
\def\nup{\nu^{\prime}}
\def\rhop{\rho^{\, \prime}}
\def\bb{b \bar{b}}
\def\cc{c \bar{c}}
\def\qq{q \bar{q}}
\def\cM{{\cal M}}
\def\cO{{\cal O}}
\def\cK{{\cal K}}
\def \ni {\noindent}
\def \be {\begin{equation}}
\def \e {\end{equation}}
\def \bea {\begin{eqnarray}}
\def \ea {\end{eqnarray}}
\def \eps {\epsilon}
\def \si {\sigma}
\def \ga {\gamma}
\def \ka {\kappa}
\def \la {\lambda}
\def \no {\nonumber}
\def \G {{\rm g}}
\def \dd {{\rm d}}
\def \Li {{\rm Li_2}}
\def \K {k_{\bot}^2}
\def \Vec#1{\mbox {\bf #1}}
\def \Veg#1{\mbox{\boldmath $#1$}}
\def \sub {\scriptscriptstyle}
\def \ps {p\hspace{-0.43em}/}
\def \mps {p\hspace{-0.45em}/}
\def \sps {p\hspace{-0.32em}/}
\def \ns {n\hspace{-0.51em}/}
\def \mns {n\hspace{-0.53em}/}
\def \ks {k\hspace{-0.49em}/}
\def \es {\epsilon\hspace{-0.47em}/}
\def \Z#1#2{\widetilde{Z}^{#1}_{#2}}
\def \Zw{\ensuremath{ {Z^{-\frac{1}{2}}_{\scriptscriptstyle W}\, } }}
\def \Zz{\ensuremath{ {Z^{-\frac{1}{2}}_{\scriptscriptstyle Z}\, } }}
\def \Zwn{\ensuremath{ \delta Z_{\scriptscriptstyle W,\,n} \,}}
\def \Zzn{\ensuremath{ \delta Z_{\scriptscriptstyle Z,\,n} \,}}
\def \slash#1{#1 \hspace{-0.42em}/}
\newcommand{\To}[2]{\stackrel{#1}{\hbox to #2 pt{\rightarrowfill}}}

\def \an {\widehat}
\def \abs#1{|\,#1\,|}
\def \vector#1{\stackrel{\hspace{-0.45em}\longrightarrow}{#1}}
\def \pa {\partial}
\def \c {\hspace{-0.2em} \cdot}
\def\wp{\ifmmode W^+\else $W^+$\fi}
\def\wm{\ifmmode W^-\else $W^-$\fi}
\def\emm{\ifmmode e^-\else $e^-$\fi}
\def\ep{\ifmmode e^+\else $e^+$\fi}

\def\sw{\ensuremath{ \sin \theta_{\rm w}}}
\def\swto{\ensuremath{ \sin^2 \theta_{\rm w}}}
\def\swfor{\ensuremath{ \sin^4 \theta_{\rm w}}}
\def\cw{\ensuremath{ \cos \theta_{\rm w}}}
\def\cwto{\ensuremath{ \cos^2 \theta_{\rm w}}}
\def\cwfor{\ensuremath{ \cos^4 \theta_{\rm w}}}

\def\ie{{\it i.e.~}}
\def\eg{{\it e.g.~}}

\def\tw{\ensuremath{ \tan \theta_{\rm w} }}
\def\ctw{\ensuremath{ \cot \theta_{\rm w} }}
\def\twto{\ensuremath{ \tan^2 \theta_{\rm w} }}
\def\ctwto{\ensuremath{ \cot^2 \theta_{\rm w} }}

\def\mw{\ensuremath{ {M}_{\scriptscriptstyle W}} }
\def\mz{\ensuremath{ {M}_{\scriptscriptstyle Z}} }
\def\mh{\ensuremath{ {M}_{\scriptscriptstyle H}} }
\def\mn{\ensuremath{ {M}_{\scriptscriptstyle N}} }
\def\mwp{\ensuremath{ {M}_{{\scriptscriptstyle W},\,{\rm phys. }}}}
\def\mzp{\ensuremath{ {M}_{{\scriptscriptstyle Z},\,{\rm phys. }}}}

\def \mt{\ensuremath{ m_t}}

\def\B#1{\ensuremath{\delta Z^{\, (1)}_{W_T}({#1})   }}
\def\S#1{\ensuremath{\delta Z^{\, (1)}_{\phi}({#1})   }}

\def \Cw#1{\ensuremath { \delta Z^{\, (1)}_{f_{#1}}(W)}}
\def \Cz#1{\ensuremath { \delta Z^{\, (1)}_{f_{#1}}(Z)}}
\def \Cga#1{\ensuremath { \delta Z^{\, (1)}_{f_{#1}}(\ga) }}

\def\L{\ensuremath{ {\rm L}}}
\def\z#1{\ensuremath { \frac{\dd z_{#1}}{z_{#1}}  }}
\def\y#1{\ensuremath { \frac{\dd y_{#1}}{y_{#1}}  }}

\def \onshellm{\vphantom{\frac{1}{1}}_{\left|_{\scr{\,p^2=m^2}}\right.}}
\def \onshell{\vphantom{\frac{1}{1}}_{\left|_{\scr{\,k^2=\mw^2}}\right.}}
\def \onshellW{\vphantom{\frac{A^2}{A^2}}\left|_{_{\scr{\,k^2=\mw^2}}}\right.}
\def \onshellZ{\vphantom{\frac{A^2}{A^2}}\left|_{_{\scr{\,p^2=\mz^2}}}\right.}
\def \onshellA{\vphantom{\frac{A^2}{A^2}}\left|_{_{\scr{\,p^2=0}}}\right.}
\def \onshellAIR{\vphantom{\frac{A^2}{A^2}}\left|_{_{\scr{\,k^2 \to 0}}}\right.}

\def \onshellH{\vphantom{\frac{A^2}{A^2}}\left|_{_{\scr{\,k^2=\mh^2}}}\right.}
\def \onshellf{\vphantom{\frac{A^2}{A^2}}\left|_{_{\scr{\,p^2=m_f^2}}}\right.}
\def \onshellp{\vphantom{\frac{1}{1}}_{\left|_{\scr{\,k^2=\mwp^2}}\right.}}
\def \onshellWp{\vphantom{\frac{A^2}{A^2}}\left|_{_{\scr{\,k^2=\mwp^2}}}\right.}

\def \dd#1{\frac{\partial}{\partial \, {#1}^2}}

\def \h#1{ \hspace*{#1mm}}
\def \v#1{ \vspace*{#1mm}}

\def\dis#1{\ensuremath { {\displaystyle  #1}}}
\def\scr#1{\ensuremath { { \scriptstyle #1}}}
\def\sscr#1{\ensuremath { { \scriptscriptstyle #1}}}

\def \myto#1#2{\ensuremath {\begin{array}[H]{c}
 {\scriptstyle {\rm #1}} \\[-2mm] {-\!\!\!-\!\!\!-\!\!\!-\!\!\!\longrightarrow} \\[-2mm] {\scriptstyle {\rm #2}}
\end{array} }}

\def \myk{\ensuremath {\begin{array}[H]{c} \\[-7mm] {-\!\!\!-\!\!\!\longrightarrow} \\[-2mm] k
\end{array} }}

\def \Myto#1{\ensuremath {\begin{array}[H]{c}
 {\scriptstyle { #1}} \\[-2mm] {-\!\!\!\longrightarrow} \end{array} }}
\def \Myeq#1{\ensuremath {\begin{array}[H]{c}
 {\scriptstyle { #1}} \\[-1mm] {=\!\!\!=\!\!\!=\!\!\!} \end{array} }}

\def \pso {p\hspace{-0.43em}/ _1}
\def \pst {p\hspace{-0.43em}/ _2}
\def \ksp {k\hspace{-0.49em}/_+}
\def \ksm {k\hspace{-0.49em}/_-}
\def \kso {k\hspace{-0.49em}/_1}
\def \kst {k\hspace{-0.49em}/_2}

\def \uo{\ensuremath { u(p_1) \, }}
\def \ut{\ensuremath { \bar{u}(-p_2) \, }}
\def \vo{\ensuremath { v(-p_3) \, }}
\def \vt{\ensuremath { \bar{v}(p_4) \, }}

\def\m{\ensuremath}

\def \co#1{\ensuremath{ [ (k - q_{#1})^2 ]}}
\def \c#1{\ensuremath{ [ (k - q_{#1})^2 - m_{#1}^2 ]}}
\def \ck#1{\ensuremath{ [ k^2 - m_{#1}^2 ]}}

\def \FP#1#2{\ensuremath { \frac{ i}{[\,{#1} ({#2})-m+i\,\eps\,]} \,\,}}
\def \PP#1#2#3{\ensuremath { \frac{(- i \h{0.5})\,  \G_{#1 #2}}{[\,{#3}^2+i\eps\,]} \,\,}}
\def \BP#1#2#3#4{\ensuremath {\,\frac{ (-i\,e) \,  \G_{#1 #2}}{[\,{#3}^2-{#4}^2+i\,\eps\,]} \,\, }}

\def \Ve#1{\ensuremath { (-i\,e \, Q_e \, \ga^{#1} ) \,\, }}
\def \Vt#1{\ensuremath { (-i\,e \, Q_t \,  \ga^{#1} ) \,\, }}

\def \Vee#1{\ensuremath { (i\,e \, v_e \, \ga^{#1} ) \,\, }}
\def \Vtt#1{\ensuremath { (i\,e \, v_t \,  \ga^{#1} ) \,\, }}

\def \Vze#1{\ensuremath { \big(i\,e \, (v_e + a_e\, \ga_5)\,
\ga^{#1} \big) \,\, }}
\def \Vzt#1{\ensuremath { \big(i\,e \, (v_t + a_t\, \ga_5)\, \ga^{#1} \big) \,\, }}
\def \Vzeo#1{\ensuremath { \big(i\,e \, (v_e^0 + a_e^0\, \ga_5)\,
\ga^{#1} \big) \,\, }}
\def \Vzto#1{\ensuremath { \big(i\,e \, (v_t^0 + a_t^0\, \ga_5)\,
\ga^{#1} \big) \,\, }}

\def \Vzere#1{\ensuremath { \big(i\,e \, (\tilde{v_e})\,
\ga^{#1} \big) \,\, }}
\def \Vztre#1{\ensuremath { \big(i\,e \, (\tilde{v_t})\,
\ga^{#1} \big) \,\, }}

\def \Vw#1{\ensuremath { \left( \frac{i\,e }{ \sqrt{2} \, \sw} \, \ga^{#1} \right) \,\, }}

\def \BD#1#2{\ensuremath {\frac{(-i)}{[\,{#1}^2-{#2}^2+i\,\eps\,]} \,\, }}


\def \Vpww#1#2#3#4#5#6#7{\ensuremath { ({#1} i\,e)\, \Big[
    (-{#6}+{#7})^{\,#2} \G^{\,{#3}\,{#4}} -   ({#7}+{#5})^{\,#3} \,
    \G^{\,{#2}\, {#4}} + ({#5}+{#6})^{\,#4} \, \G^{\,{#2} \, {#3}} \Big]\,\,  }}

\def \Vzww#1#2#3#4#5#6#7{\ensuremath { \left(\frac{{#1} -  i\,e \, \cw}{\sw}\right)\, \Big[
    (-{#6}+{#7})^{\,#2} \G^{\,{#3}\,{#4}} -   ({#7}+{#5})^{\,#3} \,
    \G^{\,{#2}\, {#4}} +  ({#5}+{#6})^{\,#4} \, \G^{\,{#2} \, {#3}} \Big]\,\,  }}

\def \e#1#2{\ensuremath { \eps_{\,#1}^{\,*}\,({#2})\,\, }}

\def \BPCu#1#2#3#4{\ensuremath {\frac{(-i)}{[\,{#3}^2-{#4}^2+i\,\eps\,]}\,
    \left( \, \G^{\, {#1}\, {#2}} + \frac{{#3}_{#1}\,
        {#3}_{#2}}{\vec{#3}^2} - \frac{{#3}_0\, [{#3}_{#1}\,
        n_{#2}+{#3}_{#2}\,n_{#1}\, ]}{\vec{#3}^2} \right)  \,\, }}

\def \eett {\ensuremath {e^+e^- \to t \bar{t} \,\,}}
\def \GeV {\ensuremath  \,{\rm GeV} \,\,}

\def\sig{\left[\frac{\displaystyle{\mathrm{d}\sigma}}{\displaystyle{\mathrm{d}\cos \, \theta}}\right]}

\def\unity{{\rm 1\mskip-4.25mu l}}
\def\re{\mathop{\mathrm{Re}}}

\newcommand{\eqir}{\stackrel{{ \rm IR }}{\longrightarrow}}

\renewcommand{\thefootnote}{\fnsymbol{footnote}}

\texttt{
\begin{flushleft}
DESY  10-144     
\\
HEPTOOLS 10-024
\\
SFB/CPP-10-79
\end{flushleft}
}
\vspace{2cm}

\bigskip

\begin{center}
{\LARGE \bf
Numerical Evaluation
of  Tensor  Feynman Integrals
\\[3mm]
in Euclidean Kinematics
}
\\
\vspace{1.5cm}
{
{\Large J. Gluza}${}^{a}$,~~
{\Large K. Kajda}${}^{a}$},~~
{\Large T. Riemann}${}^{b}$~~ and
~~{\Large V. Yundin}${}^{b}$
\vspace{1cm}

{
{${}^{a}$ Institute of Physics, University of Silesia, Uniwersytecka 4, 40007 Katowice, Poland\\ }
\smallskip
{${}^{b}$~Deutsches Elektronen-Synchrotron, DESY, Platanenallee
  6, 15738 Zeuthen, Germany \\ }
}
\end{center}

\vspace{1cm}

\begin{center}
{\Large \bf
{Abstract}}
\end{center}
For the investigation of higher order Feynman integrals, potentially with tensor structure, it is highly desirable to have numerical methods and automated tools for dedicated, but sufficiently 'simple' numerical approaches.
We elaborate two algorithms for this purpose which may be applied in the Euclidean kinematical region and in $d=4-2\epsilon$ dimensions.
One method uses Mellin-Barnes representations for the Feynman parameter representation of multi-loop Feynman integrals with arbitrary tensor rank.
Our \math{} package \ar{} has been extended for that purpose, and together with the packages \mb{} (M. Czakon) or \mbres{}
(A. V. Smirnov and V. A. Smirnov) one may perform  automatically a numerical evaluation of planar
tensor Feynman integrals.
Alternatively, one may apply sector decomposition to planar and non-planar  multi-loop  $\epsilon$-expanded Feynman integrals with arbitrary tensor rank.
We automatized the preparations of Feynman integrals for an immediate application of the package \secdec{} (C. Bogner and S. Weinzierl) so that one has to give only a proper definition of propagators and numerators.
The efficiency of the two implementations, based on Mellin-Barnes representations and sector decompositions, is compared.
The computational packages are publicly available.

\setcounter{footnote}{0}
\renewcommand{\thefootnote}{\arabic{footnote}}

\thispagestyle{empty}

\clearpage
\tableofcontents
\clearpage

\section{Introduction}
\label{sec-intro}
One goal of present calculations in particle physics is reaching higher and higher precision in perturbation theory.
Since Feynman's time  we have rules allowing to build automatically the necessary mathematical objects.
For a long term these were just the Feynman diagrams, but recently other approaches like the unitarity based perturbative approach get rising attention.
The problem remains how to evaluate the complicated integrals originating in the perturbative picture of elementary particle interactions.
Physics predicts these integrals to be defined in Minkowskian space-time.
The integrals are multi-dimensional complex functions with a complicated singularity structure.
Further, they have to be regularized by notions like $d$-dimensional space-time.
In recent years ambitious projects appeared where techniques are used to calculate physical processes in highly automatized ways \cite{Bern:2008ef,Ossola:2007ax,Moretti:2008jj}.

Here we focus on the calculation of Feynman integrals in \emph{Euclidean space-time}.
Though they are, in general, not the ultimate physical objects, their knowledge is very useful.
First, if we know them analytically, analytic continuation gives a way to transform them to the Minkowskian region, if needed so.
Moreover, if we solve, somehow, Feynman integrals involved in a given physical process analytically, we can use the knowledge in the Euclidean region to check these solutions numerically.
This has been done in numerous cases at the 2-loop level (e.g. massive Bhabha scattering \cite{Heinrich:2004iq,Czakon:2004tg,Czakon:2004wm,Czakon:2006pa,Actis:2007fs,Actis:2008br}, QCD calculations \cite{Bonciani:2008az,Beneke:2008ei,Kiyo:2008mh}),
but also at higher loop levels or in more general context \cite{Bern:2006ew,Smirnov:2010hd,Gluza:2010ws,Kosower:2010yk}).
For structural studies of quantum field theory,
the numerical calculation of Feynman integrals in Euclidean space-time has also been proven to be useful, e.g. for the check of some conjectures in super-Yang-Mills theories \cite{Bern:2006vw,Drummond:2006rz}.

In the present article we describe two publicly available computational tools based on \math{} which
facilitate numerical calculations of the kind described.

The first tool is the extended version v.2.0 of the \ar{} program \cite{Gluza:2007rt}
which generates Mellin-Barnes (MB) representations for Feynman integrals.
We discuss the construction of MB-integrals with numerators of arbitrary rank $R$ for
$L$-loop cases.
We also shortly report on additional features of older versions (v.1.1 and v.1.2)
which were released  after  publication of \cite{ambre:2010}.
We explicitely work out a variety of non-trivial numerical examples.
For this purpose, \ar{} is being  combined with the \math{} packages \mb \cite{Czakon:2005rk} and  \mbres{} \cite{Smirnov:2009up}.

The second tool is the  \math{} interface  \cs{} to the \ginac{} package \secdec{} \cite{Bogner:2007cr}.
The package \secdec{} uses the sector decomposition method to calculate general polynomial structures which are present in calculations of Feynman integrals.
In the spirit of \ar{}, we perform in \cs{} for a given Feynman integral
the automatic calculation of the characteristic $F$ and $U$ polynomials, add some normalizing factors consisting of Gamma functions and, finally, use a general formula for multi-loop tensorial Feynman
parameterisations.
The result is a
user-friendly interface to \secdec{} for the specific purpose of {tensor} Feynman integral calculations.
What remains to be done by the user
of \cs/\secdec{} is writing in a proper way the definitions of propagators
(plus numerators, if they are present).
Further,
  optional algorithmic strategies have to be chosen which are part of the core program \cite{Bogner:2007cr}.
This is the stage of automatization reached also with \ar.

The article is organized as follows.
In Section \ref{sec-def} we prepare expressions for the general multi-loop Feynman integral.
Their evaluation based on Mellin-Barnes representations with \ar{} is described in Section \ref{sec-mbnum}.
In Section \ref{sec-dec} some
details on sector decomposition and of using \cs{} are discussed,
and Section \ref{sec-comprem}
contains  numerical examples and few  comparisons of the two approaches.
It follows the Summary.
In the appendix  we list the most important  \math{} functions of \ar{} and options for \cs.

\section{Definitions}
\label{sec-def}
Comprehensive overviews of the presentations of Feynman integrals may be found e.g. in  \cite{Smirnov:book4,Heinrich:2008si}.
Here, we repeat some basic formulae in order to define our notations.
The
$L$-loop Feynman integral
in $d=4-2\eps$ dimensions with $N$ internal lines with  momenta $q_i$ and masses  $m_i$,  and $E$ external legs with momenta $p_e$ is defined here as follows:
\bea
\label{eq-bha}
G_L[T_R(k)]
=
\frac{1}{(i\pi^{d/2})^L} \int \frac{d^dk_1 \ldots d^dk_L~~T_R(k)}
     {(q_1^2-m_1^2)^{\nu_1} \ldots (q_i^2-m_i^2)^{\nu_i} \ldots
       (q_N^2-m_N^2)^{\nu_N}  }  .
\eea
The numerator $T_R(k)$ is a tensor of rank $R$ in the integration variables:
\bea\label{eq-T}
T_R(k) &=& k_{l_1}^{\mu_1}  \cdots k_{l_R}^{\mu_R} ,
\eea
and
\bea
\label{eq-num}
D_i &=& q_i^2-m_i^2 ~=~  \left[ \sum_{l=1}^{L} \alpha_{il} k_l - P_i \right]^2 - m_i^2,
\\\label{eq-mom}
 P_i &=& \sum_{e=1}^{E} \beta_{ie} p_e  .
\eea
We allow for arbitrary indices $\nu_i$, the powers of propagator functions $D_i$.

Next, the momentum integrals are replaced by Feynman parameter integrals:
\begin{eqnarray}
\label{feyn_general_param}
G_L[T_R(k)]&=&
  \frac{(-1)^{N_{\nu}}}
  {\Gamma (\nu_{1}) \dots \Gamma (\nu_{N})}
  \int
  \prod_{i=1}^{N}dx_i x_{i}^{\nu_{i}-1}
  \delta (1-\sum_{i=1}^N x_i)
~ \frac{U^{N_{\nu}-\frac{d}{2}(L+1)-R}}{ F^{N_{\nu}-\frac{d}{2}L} }
\nonumber\\
&& \times~
{ \sum_{r=0}^R}
\frac{1}{(-2)^{\frac{r}{2}}}
\Gamma\left(N_{\nu}-dL/2-r/2\right) 
~ F^{\frac{r}{2}}
{
\left\{{\cal A}_r^{[\mu_1,\ldots,\mu_r} {\cal P}_{R-r}^{\mu_{r+1},\ldots,\mu_R]}\right\}
}
 ,
\end{eqnarray}
where  $N_{\nu} = \sum_{i=1}^{N} \nu_i$.

The two functions $U$ and $F$ are characteristics of the topology of the Feynman integral.
One may derive them from
\bea
\label{bh-6}
{\mathcal N} &=& \sum_{i=1}^{N} x_i ~ D_i ~\equiv~ k(M_L)k - 2kQ + J,
\eea
where
\bea \label{eq-M}
(M_L)_{ll'}  &= & \sum_{i=1}^N  \alpha_{il}  \alpha_{il'} x_{i} ,
\\ \label{eq-Q}
Q_l  &= & \sum_{i=1}^N \alpha_{il} P_i x_i ,
\\\label{eq-J}
J  &= & \sum_{i=1}^N (P_i^2-m_i^2)x_i ;
\eea
namely:
\bea\label{eq:U}
U_L{(x)} &=& \det ( M_L),
\\\label{eq:F}
F_L({x}) &=&
-\det ( M_L)~J + Q \tilde{M_L} Q,
\eea
with
\bea\label{eq-mtilde}
\tilde{M_L} = \det ( M_L) ~ (M_L)^{-1} .
\eea
The object ${\cal A}_r {\cal P}_{R-r}$ contains the  tensor structure due to its two elements:
\bea\label{eq-a0}
{\cal A}_0&=&1,
\\\label{eq-ar}
{\cal A}_r&=&0 \mathrm{~~~for~odd~} r>0,
\\\label{eq-ar2}
{\cal A}_r^{\mu_{1} \cdots \mu_{r}}
&=&
\tilde{g}^{\mu_{1} \mu_{2}} \cdots \tilde{g}^{\mu_{r-1} \mu_{r}},
\eea
and
\bea\label{eq-pp0}
{\cal P}_{0} &=& 1,
\\\label{eq-pp}
{\cal P}_{r}^{\mu_{1} \cdots \mu_{r}}
&=&
{\cal P}^{\mu_1} \cdots {\cal P}^{\mu_r},
\eea
where we left out in the notations the indices related to the loop numbering, because they are fixed by \eqref{eq-T} when the Lorentz indices are defined:
\bea\label{eq-tilg}
\tilde{g}^{\mu_{1} \mu_{2}} &\equiv&  \left(\tilde{M_L}^{-1}\right) _{l_{1}  l_{2}  } ~ g^{\mu_{1} \mu_{2}},
\\\label{eq-tilp}
{\cal P}^{\mu_i} &\equiv&  \sum_{l=1}^L ~ \left(\tilde{M_L}\right)_{l_{i} l } ~ Q_l^{\mu_i} .
\eea
The product
$\left\{{\cal A}_r^{[\mu_1,\ldots,\mu_r} {\cal P}_{R-r}^{\mu_{r+1},\ldots,\mu_R]}\right\}$
is completely symmetrized in its Lorentz indices; take as an example ${\cal A}_2 {\cal P}_{2}$:
\bea
\label{G2V2}
{\cal A}_2^{[\mu \nu} {\cal P}_2^{\lambda\rho]}
&=&
 {\cal A}_2^{\mu \nu}   {\cal P}_2^{\lambda\rho}
 + {\cal A}_2^{\mu \lambda}   {\cal P}_2^{\nu\rho} +
    {\cal A}_2^{\nu \lambda}   {\cal P}_2^{\mu\rho}
\nl &&
+~
{\cal A}_2^{\mu \rho}   {\cal P}_2^{\nu\lambda}+ {\cal A}_2^{\nu \rho}  {\cal P}_2^{\mu\lambda} +
     {\cal A}_2^{\lambda \rho}   {\cal P}_2^{\mu\nu} ,
\eea
and, more explicitely, to e.g. $T(k_1^{\mu_1} k_2^{\mu_2})$ correspond the terms:
\bea\label{eq-ap}
{\cal A}_0 {\cal P}_2^{\mu_1\mu_2 }
&=&
P^{\mu_1} P^{\mu_2},
\eea
and,   {with a different numerical factor (see \eqref{feyn_general_param})}:
\bea\label{eq-ap2}
 {\cal A}_2^{\mu_1\mu_2} {\cal P}_0 ^{\mu_1\mu_2}
&=&
\tilde{g}^{\mu_1 \mu_2} .
\eea

For \emph{one-loop integrals},
the rotation matrix $M_1$ in the loop momenta becomes trivial, $M_1=\tilde{M_1}=1$, and so also $U_1=\det(M_1)=1$, and $F_1(x)=-J+Q^2$.\footnote{Sometimes it is useful to rewrite   $F_1(x) \to -(\sum x_i)J+Q^2$, which agrees under the integral in \eqref{feyn_general_param} with $F_1(x)$, but is now a bi-linear function of the $x_i$.}
The
(\ref{feyn_general_param}) then becomes:
\begin{eqnarray}
\label{eq:tensor1loop}
G_1(T_R)
&=&
\frac{(-1)^{N_{\nu}}}{\prod_{i=1}^N \Gamma(\nu_i)}
\int
\prod_{i=1}^{N}dx_i x_{i}^{\nu_{i}-1} \delta(1-\sum_{j=1}^N x_j)
\sum_{r=0}^R
\frac{\Gamma\left(n-\frac{d+r}{2}\right)}{ (-2)^{\frac{r}{2}} F^{n-\frac{d+r}{2}} }
\left\{{\cal A}_r {\cal P}_{R-r}\right\}^{[\mu_1,\ldots,\mu_R]} .
\nl
\end{eqnarray}
In case of e.g. $L=1, R=2$, we get for the sum in \eqref{feyn_general_param}:
\begin{equation}\label{eq-l1r2}
T(k_1^{\mu_1} k_1^{\mu_2})
\rightarrow
\Gamma(N_\nu-d/2) \left[
Q^{\mu_1} Q^{\mu_2}
- \frac{1}{2(N_\nu-d/2-1)} ~ F ~ g^{\mu_1 \mu_2}\right] .
\end{equation}
The general expressions as well as the examples agree  with \cite{Gluza:2007rt}.
For one-loop tensors, equivalent expressions are also given in \cite{Anastasiou:2005cb}.

An important observation is that the Feynman parameter representations for arbitrary Feynman integrals
would be just dependent on polynomials in the $x_i$, i.e. be sums of monomials in the $x_i$ with integer exponents, if there were not the two types of terms
$U(x)^{A(L,N,d,R)}$ and $F(x)^{B(L,N,d,r)}$.
The functions $U(x)$ and $F(x)$ are such polynomials, but they have non-integer exponents.
While, the additional terms arising from the \emph{tensorial} structure of the $L$-loop Feynman integrals are polynomials in the $x_i$.\footnote{Because eqn. (13) of \cite{Gluza:2007rt} was not sufficiently simplified, this was not evident there for $L>1$.}
The function $U(x) = \sum_n m_n(x_i)$ depends only on monomials $m_n(x_i)$ and is positive semi-definite, while function
$F(x)= \sum_{n'} [-s_{n'}]m_{n'}(x_i)+ U(x) \sum_j^N x_j m_j^2$
 depends also on the kinematical invariants and on the masses of the problem.
For Euclidean kinematics, all the $[-s_{n'}] \geq 0$ and also  $F(x)$ becomes  positive semi-definite \cite{Heinrich:2008si}.
One typical example is the $F$-function \eqref{eq-fpol}.

The above formulae may be used for automated evaluations of specific Feynman integrals.
For that purpose, one has to develop methods for their proper treatment, and two of them are worked out here.

\section{Integrations using Mellin-Barnes representations}
\label{sec-mbnum}
{Iterated applications of Mellin-Barnes' formula}
\begin{eqnarray}\label{eq-melbar}
\frac{1}{[A(s_n) m_n(x) +B(s_{n'})  m_{n'}(x) ]^a}
&=&
\frac{1}{2 \pi i}
\int\limits_{-i \infty}^{i \infty}
d \sigma [A(s_n) m_n(x)]^\sigma
[B(s_{n'})  m_{n'}(x)]^{-a-\sigma}
\nl&&
\times ~ \frac{\Gamma{(a+\sigma)}\Gamma{(-\sigma)}}{\Gamma(a)}
\end{eqnarray}
 may be used to transform the $x$-integrand of \eqref{feyn_general_param} into a sequence of monomials in the $x_i$, allowing thus to perform the $x$-integrations applying
\begin{eqnarray}\label{eq-a1}
\int_0^1 \prod_{j=1}^N dx_j ~ x_j^{\alpha_j-1}
~ \delta\left(1-\sum_{i=1}^N x_i\right)
&=&
\frac{\prod_{i=1}^N \Gamma(\alpha_i)}
{\Gamma ( \sum_{i=1}^N \alpha_i )}
.
\end{eqnarray}
One remains with the problem to evaluate  multi-dimensional complex Mellin-Barnes integrals.

The publicly available package \ar{} prepares these Mellin-Barnes integrals.
Since the first publication of \ar{} in \cite{Gluza:2007rt}, several features has been added to the package.
In \ar{}v.1.1, the  MB-representations can be constructed only for tensor integrals where all momenta of the tensor $T(k_i)$ are {multiplied by external momenta, i.e. are part of}  scalar products.
Since \ar{}v.1.2 it is foreseen to generate MB-representations for just tensor integrals.
Additionally, some new options were added, consult for details on them the webpages \cite{ambre:2010}, where also appropriate examples are documented.
One of the options allows to generate Feynman parameter representations without performing the $x$-integrations, leaving them to be performed by the preferred technique of the user.

Here we focus on \ar{} v.2.0 which generates \emph{tensor} MB-integrals in a \emph{fully} automatic way.
Previous versions have the  option \texttt{Fauto} which allows for manual manipulations on the $F$ polynomial.
Sometimes this is useful and helps to obtain a smaller dimensionality of MB-integrals (see e.g. \cite{Gluza:2007rt} and the discussion for pentagons there).
However, for many real processes with a large number of amplitudes, complete  automatization is necessary.
A second goal of the present version is a construction of MB-representations for tensorial planar $n$-loop tensors.
Though non-planar topologies can be also
obtained directly with the \emph{loop-by-loop method} employed in \ar{} \cite{Gluza:2009mj},
results may come out wrong.\footnote{We applied \ar{} successfully to e.g. the massless 2-loop vertex first studied in \cite{Gonsalves:1983nq}, which is a one-scale problem.
A typical multi-scale problem is discussed in Section 3.2 of \cite{Czakon:2007wk}.}
We recommend not to use \ar{} {without independent checks} for non-planar problems with several kinematical scales;
they hopefully shall be treated {more properly} in the future.

\begin{figure}[t]
  \begin{center}
\scalebox{0.25}{\includegraphics{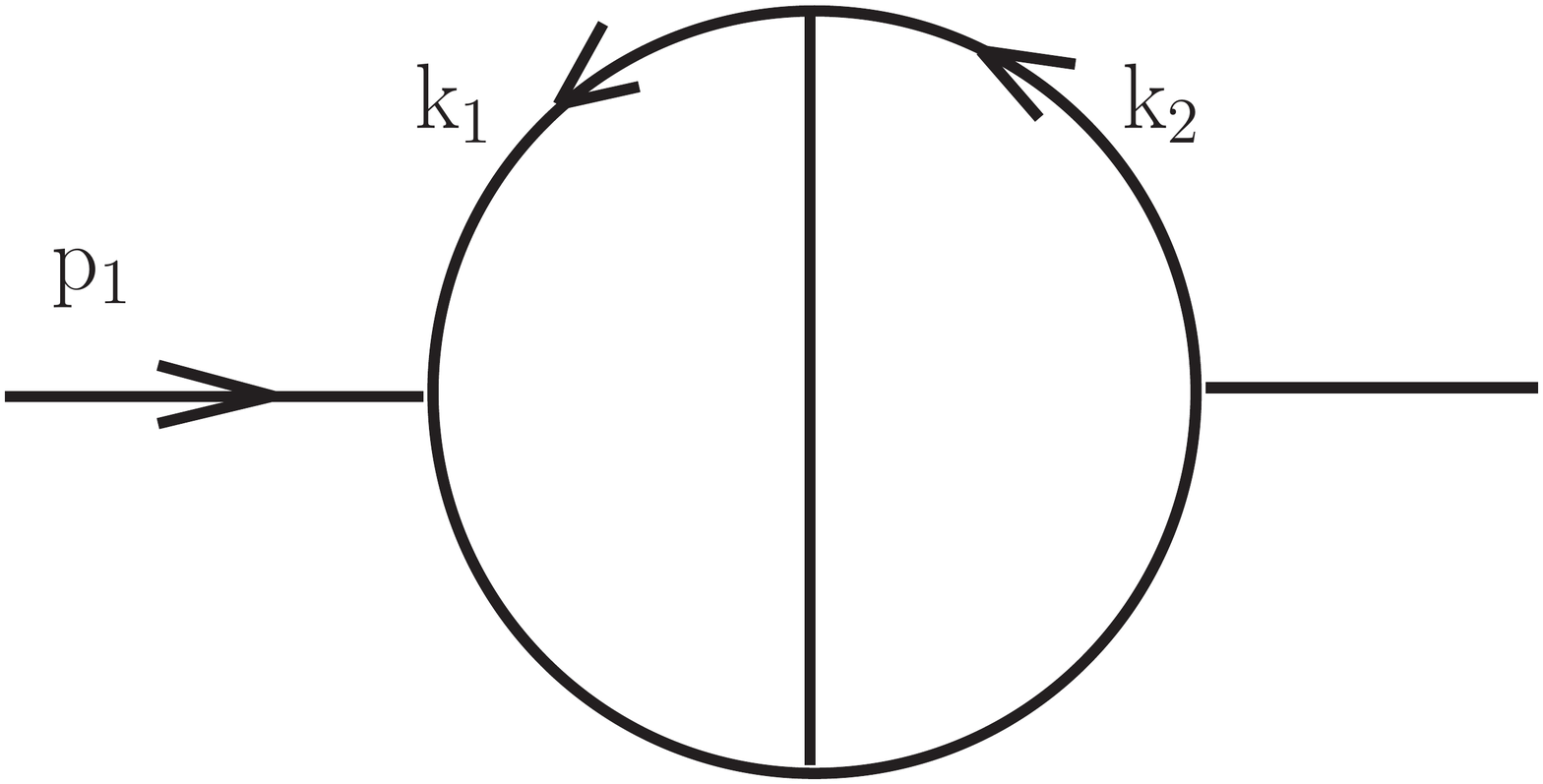}}
  \end{center}
  \caption{A simple example for the discussion of the tensor algorithm of \ar.}
\label{SEexample}
\end{figure}

In order to include tensor structures of Feynman integrals, we had to  modify properly the
iterative procedure of \cite{Gluza:2007rt}.
The best way to explain this might be an explicit example, see Figure \ref{SEexample}.

Here all propagators are assumed to be massless and the numerator is of rank $R=3$:
\begin{equation}
        \int
        \frac{(k_1\cdot p) (k_1\cdot p) (k_2 \cdot p)}
        {[k_1^2]^{n_1} [(k_2-k_1)^2]^{n_2} [(k_1+p)^2]^{n_3} [k_2^2]^{n_4} [(k_2+p)^2]^{n_5}}
        d^d k_1 d^d k_2.
 \label{SE-2loop-tensor}
\end{equation}

The calculation starts by working out the sub-loop integration over $k_1$, which leads to the following $F$ polynomial:
\begin{equation}
\label{eq-fpol}
        F=[-k_2^2] ~ x_1 x_2+ [-s] ~ x_1 x_3 + [- (k_2 + p)^2] ~ x_2 x_3 .
\end{equation}
According to (\ref{eq:tensor1loop}), the generated tensor structure
for the first sub-loop is an expression with five
separate parts; we omit here a normalizing factor and end up with a table of five different integrals:
\begin{eqnarray}
        P^{\mu_1} P^{\mu_2} \oplus \tilde{g}^{\mu_1 \mu_2}
        &\rightarrow&
        Q^{\mu_1} Q^{\mu_2} \oplus g^{\mu_1 \mu_2}
\nonumber\\
        &\rightarrow&
        (k_2^{\mu_1} x_2 - p^{\mu_1} x_3)(k_2^{\mu_2} x_2 - p^{\mu_2} x_3)
\oplus g^{\mu_1 \mu_2}
\nonumber\\
        &\rightarrow&
        \{
                k_2^{\mu_1}k_2^{\mu_2}x_2^2,
          ~ -k_2^{\mu_2}p^{\mu_1}x_2 x_3,
          ~ k_2^{\mu_1}p^{\mu_2}x_2 x_3,
          ~ p^{\mu_1}p^{\mu_2}x_3^2,
          ~ g^{\mu_1 \mu_2}
        \} .
 \label{tensor-fragmentation-schemat}
\end{eqnarray}
Evidently, we will have to perform the second sub-loop
integration over $k_2$ separately for all the above parts because they have different tensor ranks.
As a rule, the rank of a given integral in the next step will include higher rank tensors than the original one.
The situation after the first momentum integration can be symbolized
as follows:
\begin{eqnarray}\label{eq-su}
        &&
        \int
        \frac{ d^d k_2~
        p_{1\mu_1} p_{1\mu_2} (k_2 \cdot p)}
        {
                [k_2^2]^{-z_1}
                [(k_2+p)^2]^{-3+\epsilon + n_1 + n_2 + n_3 + z_1 + z_2}
        }
\nonumber\\
        &&\times~
        \{
                k_2^{\mu_1}k_2^{\mu_2} \times \text{MB}_1,
          ~ -k_2^{\mu_2}p^{\mu_1} \times \text{MB}_2,
          ~   k_2^{\mu_1}p^{\mu_2} \times \text{MB}_3,
          ~   p^{\mu_1}p^{\mu_2} \times \text{MB}_4,
          ~   g^{\mu_1 \mu_2} \times \text{MB}_5
        \}
        ,
\nl
\end{eqnarray}
where the expressions $\text{MB}_i$ stand for different  Mellin-Barnes parts of the net integral.
These parts are different  due to different Feynman parameters in \eqref{tensor-fragmentation-schemat}.
As one explicite example we reproduce $\text{MB}_1$:
\begin{eqnarray}\label{eq-subn}
        \text{MB}_1
        &=&
        (-1)^{2-\epsilon-z_2} (-s)^{z_2}
        \Gamma(2-\epsilon-n_1-n_2-z_1)
        \Gamma(-z_1)
        \Gamma(4-\epsilon-n_1-n_3-z_2)
\nonumber\\
        &&\times~
        \Gamma(-z_2)
        \Gamma(n_1+z_1+z_2)
        \Gamma(-2+\epsilon+n_1+n_2+n_3+z_1+z_2)/
\nonumber\\
        &&
        \left[ \Gamma(n_1) \Gamma(n_2) \Gamma(6-2 \epsilon-n_1-n_2-n_3) \Gamma(n_3)\right] .
\end{eqnarray}
Working with more  than two loops, additional iterations can produce further 'fragmentations' of  the expression.

In the program, all above steps are hidden to the user. The only action is to define the input object:
\begin{verbatim}
 invariants = {p1^2->s};
 MBrepr[{k1*p1,k1*p1,k2*p1},{PR[k1,0,n1]*PR[k2,0,n2]*
        PR[k2-k1,0,n3]*PR[k1+p1,0,n4]*PR[k2+p1,0,n5]},{k1,k2}]
\end{verbatim}
The output is:
\begin{verbatim}
	{-(((-1)^(n1+n2+n3+n4+n5)*(-s)^(4-2*eps-n1-n2-n3-n4-n5)*s^3
	Gamma[2-eps-n1-n3-z1]*Gamma[-z1]*Gamma[5-eps-n2+z1]
  *Gamma[4-eps-n1-n4-z2]*Gamma[4-2*eps-n1-n3-n4-n5-z1-z2]
  *Gamma[-z2]*Gamma[-4+2*eps+n1+n2+n3+n4+n5+z2]*Gamma[n1+z1+z2]
  *Gamma[-2+eps+n1+n3+n4+z1+z2])/(Gamma[n1]*Gamma[n3]
  *Gamma[6-2*eps-n1-n3-n4]*Gamma[n4]*Gamma[n2-z1]
  *Gamma[9-3*eps-n1-n2-n3-n4-n5-z2]
  *Gamma[-2+eps+n1+n3+n4+n5+z1+z2])),...}
\end{verbatim}
\begin{equation}\label{reprSE5l0m}
\end{equation}
The dots stand for the remaining four MB-integrals.
The complete expression can be found in the file  \verb+MB_SE5l0m.nb+ at the webpage \cite{ambre:2010}.

\section{Integrations by sector decomposition}
\label{sec-dec}
{The second approach --  performing sector decompositions --  transforms the $x$-integrand into a sequence of expressions where the singularities at $d=4$, as a function of $\eps$ ($d=4-2\eps$), are separated such that the arising expressions can be smoothly integrated.}
An appropriate algorithm for the automated computation of the $\epsilon$ series
of multi-loop integrals has been formulated in \cite{Binoth:2000ps,Binoth:2003ak}.
In \cite{Bogner:2007cr}, a publicly available  program is described which calculates the Laurent expansion of the following type of parametric integrals in $\eps$:
\begin{equation}
 \label{basic_integral}
 \int\limits_{x_j \ge 0} d^nx \;\delta(1-\sum_{i=1}^n x_i)
 \left( \prod\limits_{i=1}^n x_i^{a_i+\eps b_i} \right)
 \prod\limits_{j=1}^r \left[ P_j(x) \right]^{c_j+\eps d_j}.
\end{equation}

The $a_i, b_i, c_j$ and $d_j$ are integers,
and the $P_j$ are polynomials in the variables $x_1$, ..., $x_n$.
The program may handle a product of several polynomials,  and it is not required
that the polynomials are homogeneous.
These features are important for applying \eqref{basic_integral} to our tensorial structures:
it allows to calculate $G_L(T)$  \eqref{feyn_general_param}, and some
examples are given in \cite{Bogner:2007cr} for  \emph{scalar} integrals.

Here we present an interface which directly calculates
(\ref{feyn_general_param})
using as the backbone the already programmed structure (\ref{basic_integral}).
What is expected  by interface from the user is a proper definition
of the integral (\ref{eq-bha}) and of the kinematical invariants to be used.
Formally, it has to be done in the following way:
\begin{verbatim}
invariants = {invariants as a rules};
DoSector[{numerator},{denominator},{internal momenta}][low, des];
\end{verbatim}
\begin{equation}
 \label{FI}
\end{equation}
The first three entries in \verb+"DoSector"+ and $\verb+"invariants"+$ in (\ref{inputfile}) have the same form as in \ar{}.
The $\verb+"low"+$ is the leading singularity of the considered integral.
Sometimes it is not easy to determine this power of singularity.
In order to be save,
we propose to start with one or two powers of $\epsilon$ less than being relevant, giving
 zeroes for these cases.
The \verb+"des"+ is the highest power of $\epsilon$ to which we are going to calculate an integral.
One has to be careful because it is not possible to calculate only one $\epsilon$ term if it is not the  leading singularity:
If we want to calculate an $\epsilon^n$ term, but if lower terms in $\epsilon$ are present, we must start from the leading term.
This feature arises because the program calculates the whole Laurent series in $\epsilon$, taking into account also the prefactor
$\frac{(-1)^{N_{\nu}}}{\prod_{i=1}^N \Gamma(\nu_i)}$
in (\ref{feyn_general_param}).

The user input for \cs{} is extremely simple. Here we reproduce a (slightly shortened) massless 3-loop vertex
example with rank $R=3$; the topology is shown in Figure \ref{fig-mercedes}:

\begin{figure}[tb]
  \begin{center}
\scalebox{0.3}{\includegraphics{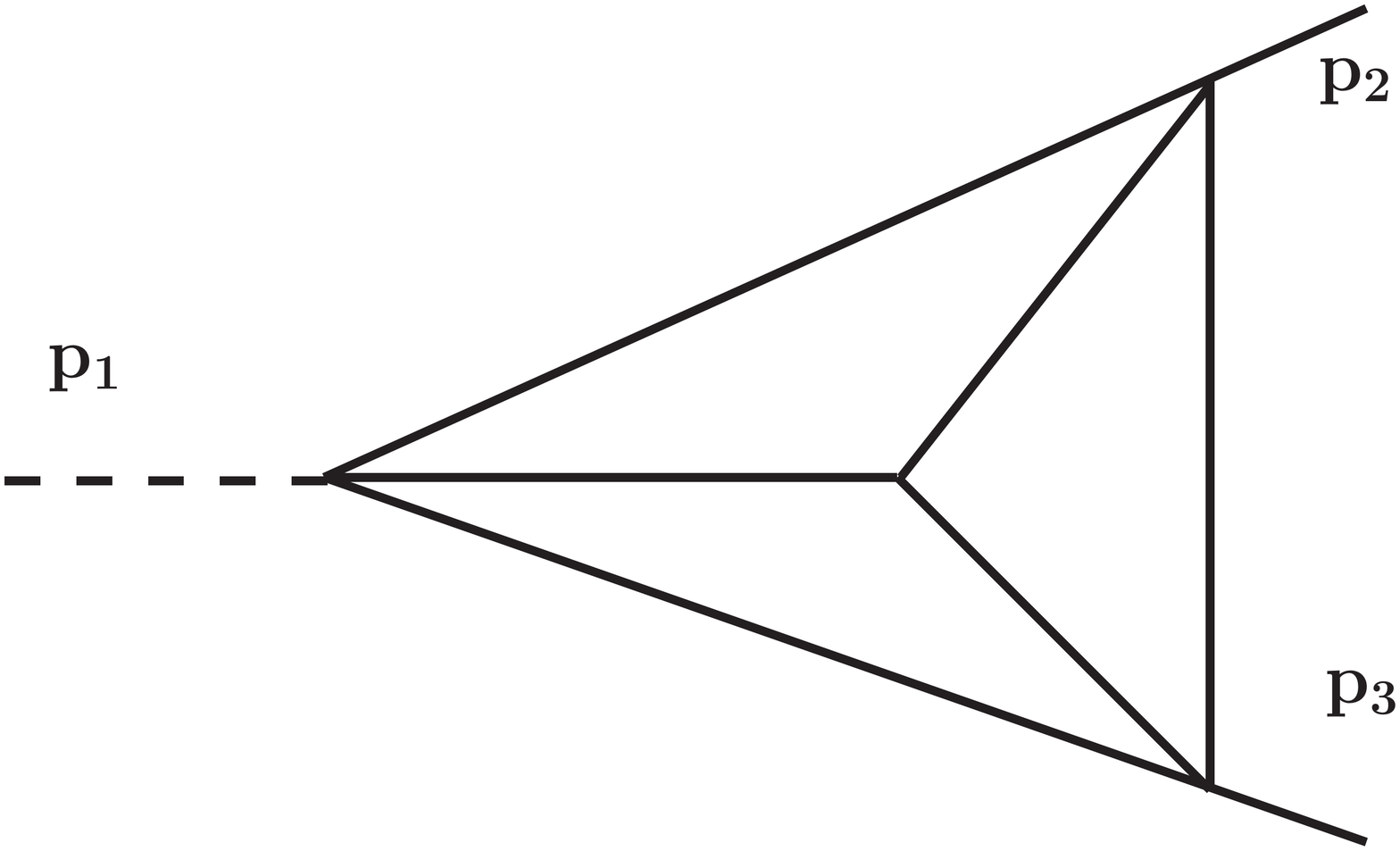}}
  \end{center}
  \caption{Topology of the three-loop vertex exemplified in the text.
}
\label{fig-mercedes}
\end{figure}

\bigskip

{\bf{\underline{ Input:}}}
\begin{equation}\label{inputfile}\end{equation}
\begin{verbatim}
************************  BEGIN  *****************************
<< CSectors.m;
x=-11;
invariants={p1*p2->1/2*x,p1*p3->1/2*x,p2*p3->1/2*x,p1^2->x,
            p2^2->0,p3^2->0};
DoSectors[{k2*p2,k3*p1,k3*p1},{PR[k2+k3+p2,0,1]*PR[k1-k3,0,1]*
          PR[k3,0,1]*PR[k1+k2-p1+p2,0,1]*PR[k1,0,1]*PR[k2,0,3]},
          {k1,k2,k3}][-5,0];
************************  END ********************************
\end{verbatim}

{\bf{ \underline{ Output:}}}
\begin{equation}\label{outputfile}\end{equation}
\begin{verbatim}
************************  BEGIN  *****************************
CSectors by K.Kajda and V.Yundin ver:1.0
last modified 22 sep 2009

Using strategy C
U & F polynomials:

   U  =  x4 (x5 x6 + x3 (x5 + x6)) + x1 (x4 x5 + x2
         (x3 + x4 + x5) + x4 x6 + x5 x6 + x3 (x5 + x6)) + x2
         ((x4 + x5) x6 + x3 (x4 + x5 + x6))
   F  = 11 (x3 (x2 + x4) + x1 (x3 + x4)) x5 x6

   Q11 = ...
   Q12 = ...
   Q21 = 121*x2*x3^2*x6^2/4
   Q22 = ...

Generating c++ source...Int11...Int12...Int21...Int22...done
Compiling source code...Int11...Int12...Int21...Int22...done
Running binary file.....Int11...Int12...Int21...Int22...done

InputForm=
{-7.622574999999999 - 0.0260407/eps^4
 + 0.049527000000000015/eps^3 - 0.4168788/eps^2 + 0.56955/eps,
 {1.3667737639753552, 2.85804370185272*^-6/eps^4,
  0.00009220935574625821/eps^3, 0.0004811810295896961/eps^2,
  0.006549529654501916/eps}}
************************  END ********************************
\end{verbatim}

The functions $F$ and $U$ in (\ref{outputfile}) are calculated with \math,
in the same way as in \ar{}.
The parameters \texttt{Q11} ... \texttt{Q22} are polynomials in the $x_i$ which arise due to the numerators in (\ref{feyn_general_param}) for tensor integrals.
These terms are select as positive semi-definite pieces \texttt{Q11} ... \texttt{Q22} in order to satisfy the conditions of the sector decomposition algorithm.
The dots for \texttt{Q11}, \texttt{Q12}, \texttt{Q22} in  (\ref{outputfile}) indicate expressions of some length and their
explicit form can be found in file \texttt{output\_mercedes} at the webpages \cite{csectors:2010}.

After \texttt{``Generating c++ source``}  and \verb+"Compiling source code"+ in (\ref{outputfile}),
the \texttt{C++} files are running.
They have the structure discussed in \cite{Bogner:2007cr}
and a user can inspect them by switching in the option
\verb+"TempFileDelete->False"+ in (\ref{FI}).
Other useful options are listed and described in the Appendix.

By default, the numerical errors of the results are also given.
They are calculated by taking the combined  error for all the integrals $I_i$ calculated at a given term of the $\epsilon$-expansion:
\begin{equation}\label{eq-erro}
 \Delta I_\epsilon = \sqrt{ \sum_{i=1}^{A} (\Delta I_i)^2}.
\end{equation}
As it is well known, some of integrals, especially massless ones, can be difficult to integrate numerically.
Here the proper choice of a sector decomposition strategy may help \cite{Bogner:2007cr}. In (\ref{outputfile}), the calculation has been done with optional strategy C.
In order to control which $\epsilon$-term is currently calculated, the default option \verb+TempFileDelete->False+ produces a log file.
For the \texttt{Q11} term of the example, it is:

\begin{verbatim}
 >> SE5l0m_num11.log <<
Order eps^(-2): 0 +/- 0
Order eps^(-1): 0 +/- 0
Order 1: -203.056 +/- 0.00948256
Order eps: 153.241 +/- 0.0721185
\end{verbatim}

A list of all options can be obtained with the command  \verb#?Options# in \math{}.

In order to test \cs{} we have checked many topologies: multi-loop tadpoles and self-energies,
three-loop vertices up to rank $R=  5$ and with double dots on propagators (corresponding to setting index $\nu=3$, because a `dot' raises the index by one), four-point functions up to rank $R=  4$, and some one-loop five- and six-point functions.
For some higher rank tensors we have used Integration-by-Parts decompositions of the Feynman integrals using the computer algebra package \texttt{IdSolver} \footnote{\texttt{IdSolver} is an unpublished \texttt{C++} package performing Feynman integral reductions with the Laporta algorithm.  J.G. thanks M. Czakon for the opportunity to use it.}. Some numerical examples of these tests can be found in the file 
\verb+numerical_checks.nb+ at the webpages \cite{csectors:2010}.

\section{Numerical results}
\label{sec-comprem}
In Section \ref{sec-mbnum} it has been shown how to define Feynman integrals and how to get numerical values for them at  chosen kinematical points using
the MB-method.
For sector decomposition, the same has been discussed in Section~\ref{sec-dec}.
For scalar integrals it is straightforward to use, together with  \ar{}, the \mb{} package, and to perform the  numerical integrations \cite{Czakon:2005rk}.
For tensor integrals, especially with loop order $L>1$, we have usually many MB-integrals for which the command \verb+MBrules+ of \mb{} has to be performed in order to find proper integration paths for the MB-integrations.
Sometimes, depending on the degree of divergency of the original Feynman integral, this is not possible if only an analytical continuation in $\epsilon$ is done.
Then, an analytical continuation in one of the indices, may be successful.
This has also been automatized.
 If \verb+MBrules+ does not find a valid rule for some $\epsilon \neq 0$, then the power of the first propagator $\nu_1$ is changed, e.g.
if $\nu_1=1$, then $\nu_1 \to \nu_1=1+\eta$ is applied.
If again \verb+MBrules+ cannot find a valid rule, $\nu_2$ is treated analogously, and the procedure can be continued until \verb+MBrules+ is successful.\footnote{If it happens that an analytical continuation in only one additional parameter $\eta$ is not sufficient, the program will stop with a proper remark.}

We introduced the auxiliary file \texttt{MBnum.m} to \mbm{} which realizes this procedure
and the subsequent automatized analytic continuation, $\epsilon$-expansion,
and numerics. The file may be obtained from the webpages \cite{ambre:2010}.

If an {already prepared} MB-representation \verb+repr+
is {available}, e.g. in (\ref{reprSE5l0m}), it is enough to use the \texttt{MBnum} function (see Appendix for details), e.g.:

\begin{verbatim}
MBanalytic=MBnum[repr, -1, {s -> -11},
                {n1 -> 1, n2 -> 1, n3 -> 1, n4 -> 1,  n5 -> 1}, 2]
res=MBintegrate[MBanalytic,{s->-11}]
\end{verbatim}

This gives the following numerical result (see also file \verb+out_SE5l0m+ at the webpages  \cite{ambre:2010}):
\begin{eqnarray}\label{SEnumMB}
{\rm SEnumMB}&=&-7.5625/\epsilon^2-20.4506/\epsilon
-178.18\pm0.0171936
+(18.3642\pm 0.0248465)\; \epsilon \nonumber \\
\end{eqnarray}
The corresponding values returned by \cs{} are
(see also file \verb+output_SE5l0m+ at \cite{csectors:2010}):

\begin{verbatim}
SEnumSD=
{-178.1927 - 7.56258/eps^2 - 20.4505/eps + 18.394000000000005*eps,
{0.011130644628528934, 0.00029563/eps^2,
0.00302109/eps, 0.06629380324170578*eps}
}
\end{verbatim}
\begin{equation}\label{SEnumSD}\end{equation}

\begin{figure}[tb]
\begin{center}
\scalebox{0.25}{\includegraphics{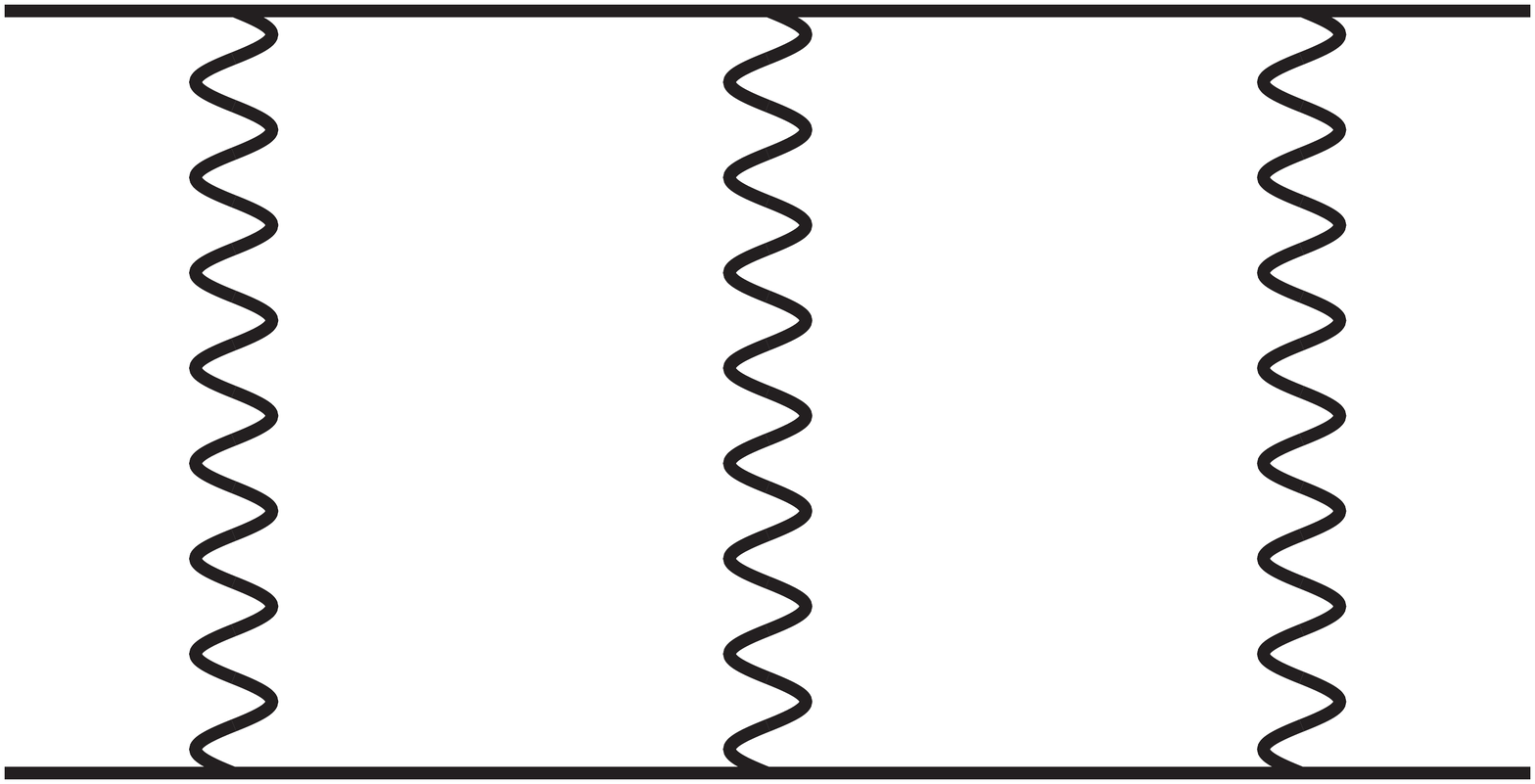}}
\vspace{-1cm}
\end{center}
\caption{The planar two-loop Bhabha topology B1.}
\label{B1fig}
\end{figure}

Another example is two-loop Bhabha scattering.
For two-loop 4-point functions, sector decomposition
 needs a lot of RAM (typically up to  few GB)
and also of computing time.
In such cases, often the numerical integrations are done faster  using the MB-method.
Some numerical results for the Bhabha Feynman integral with rank $R=2$,
\begin{eqnarray}
\textrm{B1} &=&
  \int         d^d k_1 d^d k_2
        \frac{(k_1\cdot p_1) \; (k_2 \cdot p_2)}
{(k_1^2-m^2) (k_1 + p_1)^2 [(k_1 + p_1 + p_2)^2- m^2]} \nonumber \\
&\times& \frac{1}{(k_1- k_2)^2 (k_2^2-m^2)[(k_2 + p_1 + p_2)^2- m^2]
(k_2 + p_1 + p_2 + p_4)^2} ,
\label{B1rank2}
\end{eqnarray}
corresponding to the topology shown in Figure~\ref{B1fig},
can be found in Table~\ref{topology-B1}.

\begin{table}[t]
\centering
 \begin{tabular}{|l|r@{.}l|r@{.}l|r@{.}l|r@{.}l|}
\hline
     &\multicolumn{2}{|c|}{\texttt{AMBRE} + \texttt{MB} }
     &\multicolumn{2}{|c|}{\texttt{CSectors}, X         }
     &\multicolumn{2}{|c|}{\texttt{AMBRE} + \texttt{MB} }
     &\multicolumn{2}{|c|}{\texttt{CSectors}, C         }
\\
\hline
       & \multicolumn{4}{|c|}{$m=0$}& \multicolumn{4}{|c|}{$m=1$}
\\
\hline
       $\epsilon^{0}$   &  0&65734(6) &  0&659(2)&3&186(6)  & 3&174(2)
\\
        $\epsilon^{-1}$ &  0&13921(8) &  0&1396(5) & 1&0383(1) & 1&0381(3)
\\
        $\epsilon^{-2}$ &  0&018095   &  0&01835(9)&0&28817(1) & 0&28816(4)
\\
        $\epsilon^{-3}$ &  0&104974   &  0&10500(2) &\multicolumn{2}{|c|}{--} &\multicolumn{2}{|c|}{--}
\\
        $\epsilon^{-4}$ &-- 0&0217857  &-- 0&021785(3)&\multicolumn{2}{|c|}{--}  &\multicolumn{2}{|c|}{--}
\\
\hline
        T \texttt{[$s$]}  & \multicolumn{2}{|c|}{368}
& \multicolumn{2}{|c|}{26700 ~ (7.5 $h$)}&\multicolumn{2}{|c|}{ 945 }&\multicolumn{2}{|c|}{ 70220 ~ (19.5 $h$)}
\\
\hline
        \multicolumn{9}{|c|}{$s=-5$, ~~ $t=-7$}
\\
\hline
\end{tabular}
   \caption{
Numerical values for the first terms of the $\eps$ expansion of the massless and massive Bhabha 2-loop double box with tensor rank $R=  2$, defined in \eqref{B1rank2}. The topology is shown in figure \ref{B1fig}.
The package \cs{} was used with strategy X for $m=0$ and with strategy C for $m\neq0$.}
\label{topology-B1}
\end{table}

The Mandelstam variables are $s=(p_1+p_2)^2$ and $t=(p_2+p_4)^2$.
The so-called reducible numerators of a tensor Feynman integral can be  contracted with propagators.
Apart from speeding up calculations, such contractions have been used to check
the implementation of tensor structures into \ar{} and \cs.
In our example, the relation
\begin{eqnarray}
       (k_1\cdot p_1) &=& \frac 12 \left[
(k_1 + p_1)^{-2} -
(k_1^2-m^2)^{-1}-2 m^2 \right]
\label{rel1}
\end{eqnarray}
may be used to change the rank $R=  2$ tensor of \eqref{B1rank2} into three tensor integrals of rank $R=  1$
(in addition reducing the number of propagators by one for two of them).
In this way, the numerical results given in Table~\ref{topology-B1} can be cross-checked.

\begin{figure}[t]
  \begin{center}
{\includegraphics[angle=-90,width=0.4\textwidth]{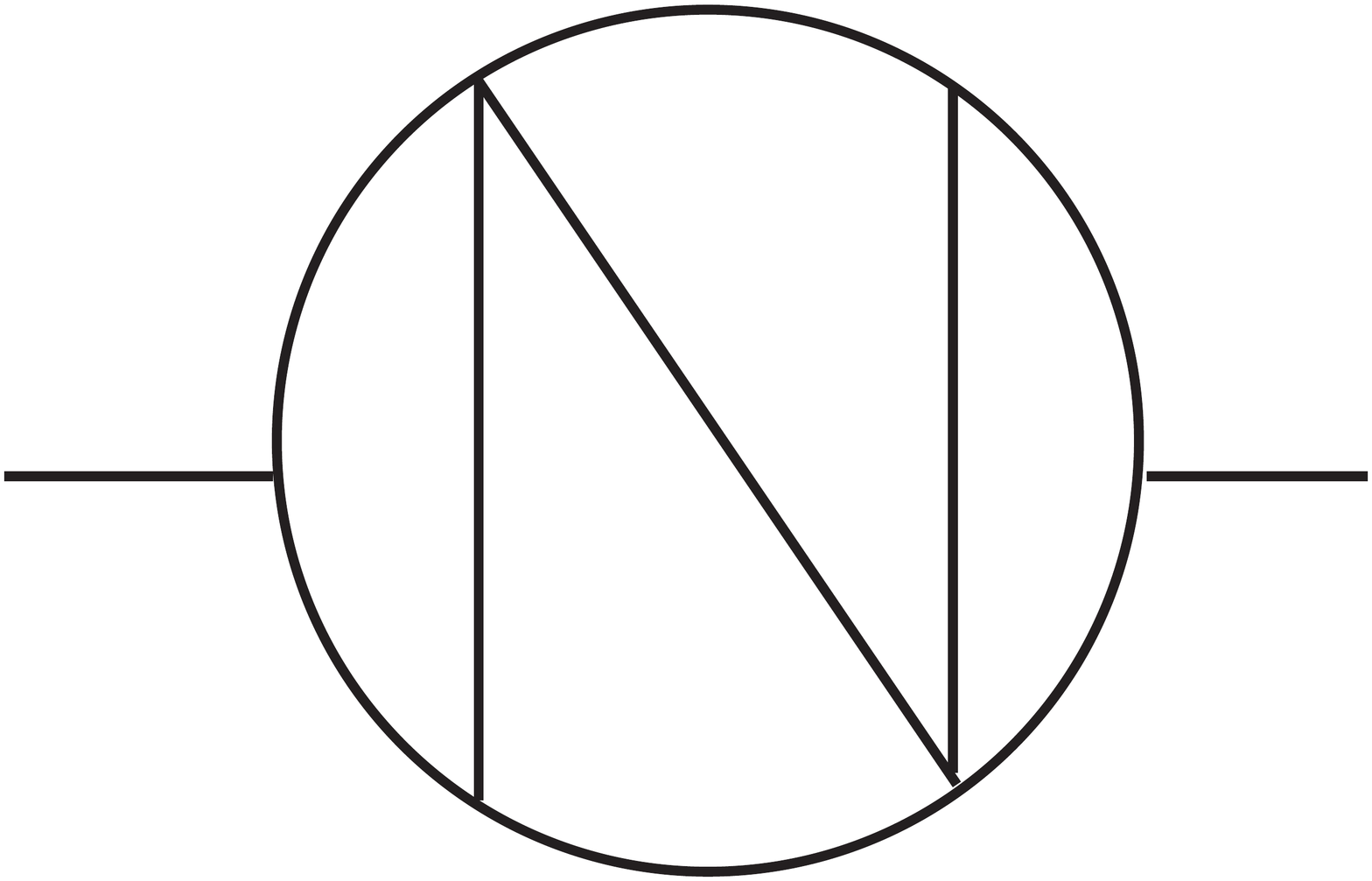}}
  \end{center}
  \caption{Four-loop self-energy. For details of its calculation see the
files \texttt{MB\_SE4loop.m}, \texttt{MB\_SE4loop.out}
and \texttt{SD\_SE4loop.sh}, \texttt{SD\_SE4loop.out} at the webpages \cite{ambre:2010,csectors:2010}.
}
\label{4SE}
\end{figure}

Further, let us present a four-loop self-energy scalar diagram,
Fig.~\ref{4SE}.
With \cs{} it takes a few hours to calculate the constant
term of the $\epsilon$-expansion.\footnote{More precisely, it takes eight ours
both with strategies C and X. This and the other examples calculated in this paper
have been run  on a Xeon personal computer.} As it is a scalar diagram, we can make direct
comparisons with \texttt{FIESTA2} \cite{Smirnov:2009pb}, which is much faster but applies to {scalar integrals} only.
For \ar/\mb{} the complete calculation takes about two minutes
only, which is comparable to using \texttt{FIESTA2}.

\begin{figure}[t]
  \begin{center}
{\includegraphics[angle=-90,width=0.4\textwidth]{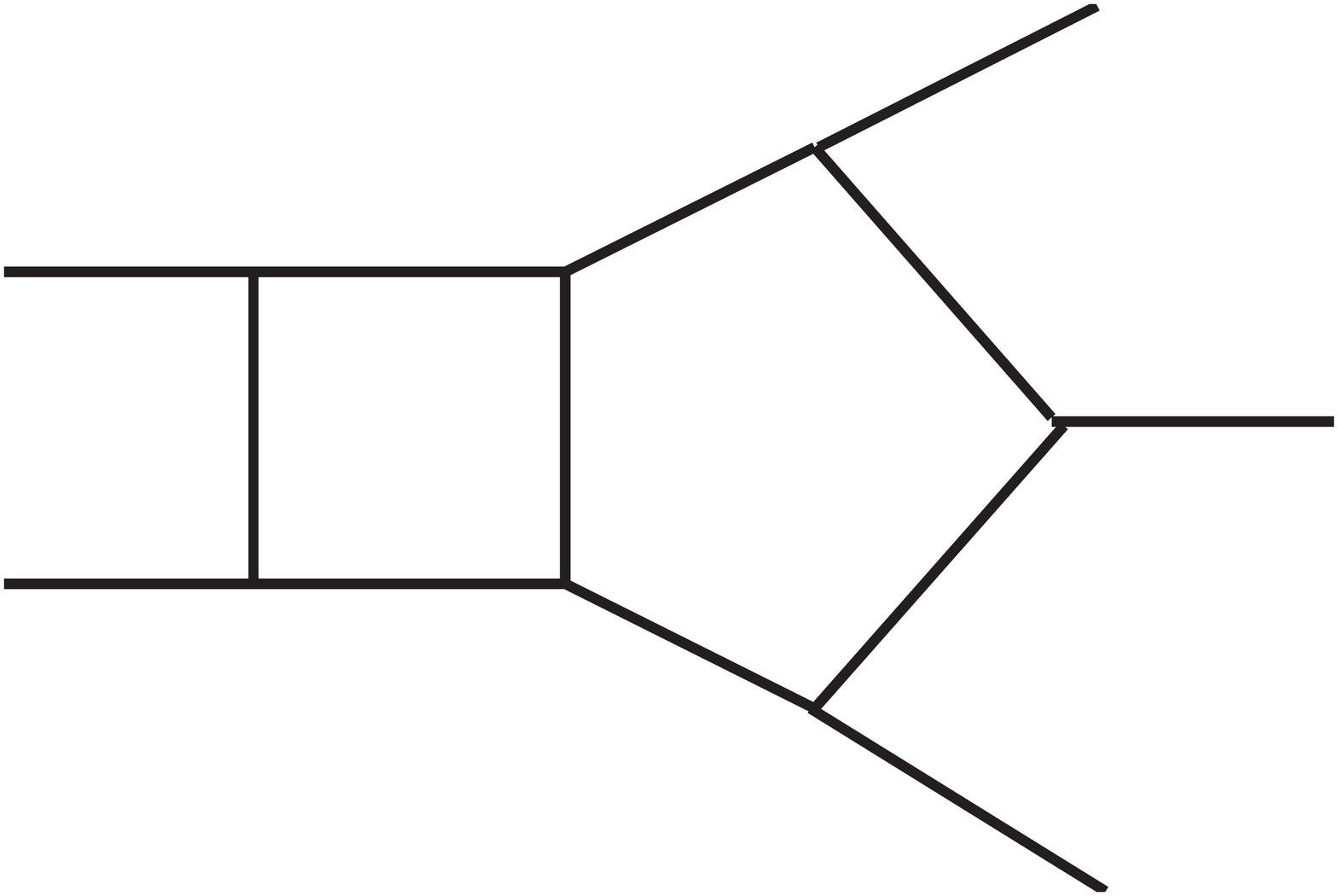}}
  \end{center}
  \caption{The pentabox topology PB.
}
\label{pbox}
\end{figure}

Finally, as a more complicated case, let us take a pentabox of rank $R=  3$:
\begin{eqnarray}
\textrm{PB} &=&
  \int         d^d l_1 d^d l_2
        \frac{(l_2\cdot k_1) \; (l_2 \cdot k_2)\; (l_1 \cdot k_5)}
{l_1^2 (l_1 - k_1)^2 (l_1 - k_1 - k_2)^2} \nonumber \\
&\times& \frac{1}{(l_1- k_1-k_2-k_3)^2 l_2^2(l_2 - k_5)^2
(l_2 - k_4-k_5)^2 (l_1+l_2)^2}
\label{pboxeq}
\end{eqnarray}
 Its topology (called PB here) is shown in Figure~\ref{pbox}.

All external momenta $k_i$ are incoming,
$s_{ij}=(k_i+k_j)^2$, $s_{12}=-1,s_{23}=-2,s_{34}=-3,
s_{45}=-4,s_{15}=-5$),
and the numerical result is:

\begin{verbatim}
PB=
{-7.431348973217098+0.328125/eps^4+0.5340786885498234/eps^3
 -0.44570891627426246/eps^2 - 3.1058689125651284/eps,
{0.012739775270198513+5.560657237081255*10^(-10)/eps^2
 +0.0003572744410650107/eps}}
\end{verbatim}
\begin{equation}\label{verb-pb}\end{equation}

Some caution must be paid to the higher dimensional MB integrals.
They
can give underestimated errors, taken from the CUBA error output.
Whether this happens can be checked switching on the MB.m option \texttt{Debug}.
The result in \eqref{verb-pb} has
been obtained for  the following set of \mb{} parameters
(see the file \texttt{MB\_PBox.m} at \cite{ambre:2010}):
\begin{verbatim}
 SetOptions[MBintegrate, Verbose->False, PrecisionGoal->10,
            AccuracyGoal->24, MaxPoints -> 1000000];
\end{verbatim}
As it can be seen in \verb+MB_PBox.m+, the
construction of the MB-representation for \eqref{pboxeq} starts with momentum $l_2$,
leading to maximally nine dimensional MB-integrals.
Applying Barnes lemmas reduces the integrals to at most seven-dimensional.
Starting with the integration over $l_1$, we get maximally eleven-dimensional integrals.
Altogether, the calculation takes about five hours: The generation of the
MB-representations needs a couple of minutes, and the rest of the time is needed for  analytic continuations and numerical integrations.

Additional instructive examples for using \ar{} and \cs{} may be found at the webpages \cite{ambre:2010,csectors:2010}.

\section{Summary}
\label{sec-summary}
The \cs{} package has been prepared as an interface to the sector decomposition  package \secdec{} for the automatic numerical evaluation of tensor Feynman integrals.
The \ar{} package has been extended for the automatic treatment of tensor structures of multi-loop Feynman  integrals.

For \cs{}, the bottleneck is the numerical evaluation based on
\verb+sector_decomposition+ and \ginac{} which, especially for
higher rank tensors and higher loop orders, consumes a huge amount of RAM.
However,
for smaller problems, up to two loops and tensors of moderate rank, the program
works well.
For more complicated cases we recommend to use the Mellin-Barnes approach.
Here there are many possibilities to optimize the way to get numerical results.
This can be done at the level of construction of MB-representations  (e.g. by a
change of order of the integration over internal momenta),
then there are different ways of analytic continuation, and finally the \math{} package
\barnesr{} \cite{mbtools} can be used to try to reduce the
dimensionality of the MB-integrals; examples for the latter can be found in \cite{Gluza:2010mz} and at \cite{ambre:2010,csectors:2010}).

For the near future it is planned to
automatize the construction of MB-representations for non-planar Feynman integrals,
what deserves to leave the loop-by-loop approach.
Further, it is foreseen to build  MB-representations for special forms of linear propagators which appear  e.g. in heavy quark effective theory (HQET) or in calculations of the QCD static potential.

\section*{Acknowledgements}
The present work is supported in part by the European Community's
Marie-Curie Research Training Networks  MRTN-CT-2006-035505 `HEPTOOLS'
and
MRTN-CT-2006-035482 `FLAVIAnet',
by the
Polish Ministry of Science and High
Education from budget for science for years 2010-2013,
grant number N N202 102638,
and by
Sonderforschungsbe\-reich/Trans\-regio 9--03 of Deutsche Forschungsgemeinschaft
`Computergest{\"u}tzte Theo\-re\-ti\-sche Teil\-chen\-phy\-sik'.

\clearpage

\section{Appendix}

The instructive examples for using \ar{} and \cs{} may be found at 
the webpages \cite{ambre:2010,csectors:2010}.

\subsection*{AMBRE (ver 1.X)}
The appropriate loop integral is defined by:\\\\
(1) {\bf a list of kinematic invariants}, e.g. 
{\bf invariants = \{p1*p1 \verb#-># s\}},\\\\
(2) {\bf Fullintegral[numerator, propagators, internal momenta, options]}.
\\
\\
{\bf invariants} must be defined before {\bf Fullintegral}.
\\
\\
The arguments of {\bf Fullintegral} are as follows:
\begin{itemize}
	\item \texttt{numerator:} numerator which can be given in the contracted form, e.g.  \verb#{k1*p1,k2*p2}# or in the uncontracted form, e.g.
\verb#{k2[mu1],k2[mu2]}#  ,
	\item \texttt{propagators:} product of propagators of the form \texttt{PR[q,m,n1]}$\equiv (q^2-m^2)^{n_1}$,
	\item \texttt{internal momenta:} list of internal momenta, e.g. 
\verb#{k1,k2}#.
\end{itemize}

This version of AMBRE uses a semi-automatic approach when building Mellin-Barnes representation. That methodology is accomplished by the following two functions.\\\\
(3) {\bf IntPart[iteration, options]} -- prepares a subintegral for a given internal momentum by collecting the related numerator, propagators and integration momentum:
\begin{itemize}
\item \texttt{iteration:} iteration for which subintegral will be prepared. In practice \texttt{IntPart} function must be executed in specific order i.e firstly \texttt{IntPart[1]} then \texttt{IntPart[2]} and so.
\item \texttt{options:}
	\begin{itemize}
		\item \texttt{Text:} it can have two boolean values \texttt{True} or \texttt{False}. Controls if additional text appears during calculations.
	\end{itemize}
\end{itemize}
(4) {\bf SubLoop[integral, options]} -- determines for the selected subintegral the $U$ and $F$ polynomials and an M-B representation.
\begin{itemize}
\item \texttt{integral:}  this argument must be left as it is.
\item \texttt{options:}
	\begin{itemize}
		\item \texttt{Text:} as in the  \texttt{IntPart} function.
		\item \texttt{Xintegration:} controls whether integration over Feynman parameters is performed or not.
	\end{itemize}
\end{itemize}

(1)-(4) constituates basic functions. An additional functions are:
\\
(5) {\bf Fauto[value]} -- allows user specified modifications of 
the $F$ polynomial \texttt{fupc}. Must be used after \texttt{IntPart} and before \texttt{SubLoop}.
\begin{itemize}
	\item \texttt{value:} can be 0 or 1. For the first one user can modify $F$ polynomial. For the latter this possibility is turned off.
\end{itemize}
(6) {\bf BarnesLemma[representation,number,options]} -- function tries 
to apply Barnes's first or second lemma to a given representation.
\begin{itemize}
	\item \texttt{representation:} M-B representation to
be checked.
	\item \texttt{number:} it has two possible values, $1$ for 
the first Barnes lemma and $2$ for the second lemma.
	\item \texttt{options:} there are the following two boolean options for this function
	\begin{itemize}
		\item \texttt{Text:} displays or does not display
an additional information.
		\item \texttt{Shifts:} searches for the pairs of two integration variables $z_i+z_j$ and $z_i-z_j$ which, after application of the appropriate shift one of it is cancelled.
	\end{itemize}
\end{itemize}
(7) {\bf ARint[result,i\_]} -- in \texttt{version 1.2}, 
it displays the MB-representation number i for Feynman integrals with numerators
\subsection*{AMBREnLOOP (ver 2.0)}

The basic functions of \texttt{AMBREnLOOP} are:\\\\
(1) {\bf MBrepr[numerator, propagators, internal momenta, options]}.
\\
It returns M-B representation for a given loop diagram. All arguments, except \texttt{options},
have the same form as parameters in \texttt{version 1.X}.
\begin{itemize}
	\item \texttt{options:}
	\begin{itemize}
		\item \texttt{Text:} displays or not information text.
		\item \texttt{OptimizedResult:} the final MB representation is written in such a way that Gamma functions are factorized. This option makes an output more condense. However sum of different Gamma functions for a given integral can cause problems if treated as it stands in analytic continuation (finding rules). Optionally it is switched off.
		\item \texttt{BarnesLemma1:}  turns on or off first Barnes lemma checking.
		\item \texttt{BarnesLemma2:}  does the same as above option but for second Barnes lemma
	\end{itemize}
\end{itemize}
Intermediate output of a given subloop is displayed in a specific format
using \verb#INT# function:\\
\verb#INT[numerator,representation,propagators1,propagators2]#\\
\begin{itemize}
\item \texttt{arguments:}
\begin{itemize}
\item \texttt{numerator:} it is, for tensor integrals,  of the form
\verb+{k2[mu1],k2[mu2]}+.
\item \texttt{representation:} Mellin-Barnes representation obtained during
the calculation of the current sub-loop/iteration part. It is multiplied by
Mellin-Barnes representations which were calculated in the previous
step/iteration.
\item \texttt{propagators1:} propagators extracted out of the $F$-polynomial in the
current iteration.
\item \texttt{propagators2:} keeps propagators in the same form as propagators1 does.
The propagators2 are independent of the present integration variable but undergo next iteration(s).
\end{itemize}
\end{itemize}
(2) {\bf BarnesLemma[representation,number,powers of propagators,options]} -- this function works as in the previous versions of \texttt{AMBRE}. The only difference is the extra parameter:
\begin{itemize}
	\item \texttt{powers of propagators:} -- here a list of powers 
of propagators which appear in an input loop integral must be given, e.g.
\verb#{n1,n2,n3}#.
\end{itemize}

\subsection*{CSectors}

The appropriate loop integral is defined by:\\\\
(1) {\bf a list of kinematic invariants}, e.g. 
{\bf invariants = \{p1*p1 \verb#-># s\}},\\\\
(2) {\bf 
DoSectors[numerator, propagators, internal momenta, options][min, max]}\\\\
The arguments of this function are exactly the same as in case of {\bf MBrepr} in the \texttt{AMBREnLOOP} package.
However, in contrast to the \texttt{AMBRE} package, 
numerator can be written only in the uncontracted form.
 
The package allows to modify its behaviour by adding additional options: 
	\begin{itemize}
		\item \texttt{SetStrategy:} chooses one of the strategies 
available in sector decomposition libraries \cite{Bogner:2007cr}.
		\item \texttt{SourceName:}  a prefixing for source,
binary and log files; the option just allows to choose any name for
the files connected with the calculation of a given integral.
		\item \texttt{TempFileDelete:} by default it is set to 
\verb#TempFileDelete->True#; when set to \verb#False#, it does not delete C++ 
source and binary files as well as log file.
		\item \texttt{LogFile:} causes \texttt{CSectors} to create 
(or not) a log file, where numerical results for given integral and epsilon 
are stored. The default is \texttt{True}
		\item \texttt{ShowErrors:} controls whether the errors 
of the numerical calculation will be displayed or not; errors are calculated 
using  the function \verb#res.get_error()# of \texttt{sector decomposition}
\cite{Bogner:2007cr}.
		\item \texttt{IterationsLow, IterationsHigh, CallsLow, CallsHigh:} these are Monte-Carlo parameters, see \cite{Bogner:2007cr} 
for a description.
		\item \texttt{compiler:} allows to choose another compiler;
the default is \texttt{g++}.
		\item \texttt{cppflags, libs:} the paths to header and 
library files, required by \texttt{sector decomposition} libraries 
\cite{Bogner:2007cr}.
	\item \texttt{min, max:} indicates minimum and maximum of the
Laurent series expansion in epsilon.
\end{itemize}
The default options can also be displayed by the command \texttt{Options[DoSectors]}.\\

\section*{}

\end{document}